\documentclass{sig-alternate}

\usepackage{amsmath}
\usepackage{oubraces}

\begin{document}
%

\title{Searching for a Unique Style in Soccer}
%
%
%
%
%

\numberofauthors{2} 
%
\author{
%
%
\alignauthor Laszlo Gyarmati, Haewoon Kwak\\
       \affaddr{{Qatar Computing Research Institute}}\\
       \email{\{lgyarmati,hkwak\}@qf.org.qa}
\alignauthor Pablo Rodriguez\\
       \affaddr{Telefonica Research}\\
       \email{pablorr@tid.es}
}

\maketitle
\begin{abstract}
	Is it possible to have a unique, recognizable style in soccer nowadays? We address this question by proposing a method to quantify the motif characteristics of soccer teams based on their pass networks. We introduce the the concept of ``flow motifs'' to characterize the statistically significant pass sequence patterns. It extends the idea of the network motifs, highly significant subgraphs that usually consists of three or four nodes. The analysis of the motifs in the pass networks allows us to compare and differentiate the styles of different teams. Although most teams tend to apply homogenous style, surprisingly, a unique strategy of soccer exists. Specifically, FC Barcelona's famous tiki-taka does not consist of uncountable random passes but rather has a precise, finely constructed structure.
\end{abstract}




\section{Introduction}
Although the basic rules of soccer have barely changed since the 1920s\footnote{the introduction of the two-opponent version of the offside rule}, they intrinsically enabled teams to develop distinctive strategies that dominated the soccer landscape for several years (e.g., the Hungarian and the Dutch team of the 50s and 70s)~\cite{jonathan2013inverting}. However, in these days the possibilities to prepare against these teams and thus ruin their strategies was limited due to technological reasons. Nowadays, technological advancements of the last decade allow team staffs to view any first division soccer game on a short notice; hence, it seems to be challenging to have and sustain a unique playing characteristic---which is additionally successful too---in the global soccer space. Does such a unique, recognizable style exist in soccer nowadays?

The identification and understanding of the style of soccer teams have practical impacts apart from the esthetics of the game. The players of a team should obey the style (i.e., strategy) of the team to maximize the team's chance to win. Hence, it is crucial to raise youngsters and to sign players who are capable of playing according to the style of the team. Failing to do so not only has an impact on the success but also on the profitability of the club. There are numerous examples of newly signed players who were not compatible with the style of their new clubs~\cite{worstsignings}, therefore, there is a need for a quantitative analysis of team's style to avoid these discrepancies.

The rareness of goals is the profoundest feature of soccer that distinguishes it from other team sports. Although the teams have 11 players and 90 minutes to score in each match, it is not unusual to have a goalless draw as the final score~\cite{anderson2013numbers}. These results are not solely due to a spectacular performance of the goalkeepers but rather a consequence of the low number of scoring chances. Hence, metrics related to scoring cannot describe the style (i.e., the strategy) of a soccer team.

Passes, on the other hand, happen numerously in every game irrespective of the quality of the teams. The pass network of a soccer team consists of the players as vertices and the passes between the players as the edges. Prior art focused either on the high-level statistics of the pass networks (e.g., betweenness, shortest paths) or the strength of the connection between pairs of players~\cite{duch2010quantifying,pena2012network,narizuka2013statistical,lucey2013assessing}. These metrics describe the static properties of a pass network, e.g., the metrics aggregate all the passes into one network, neglect the order of passes, etc. On the contrary, we focus on the dynamic aspects of the pass networks by examining the ``flow motifs'' of the teams.

We propose the concept of ``flow motifs'' to characterize the statistically significant pass sequence patterns. It extends the idea of the network motifs, highly significant subgraphs that usually consists of three or four nodes, suggested by Milo et al.~\cite{milo2002network} that mainly apply to the static complex networks (e.g., food webs, protein-structure networks, and social networks). We extend their work towards ``flow motifs'' to analyze pass networks that are highly dynamic and in which the order of connections is important. Our methodology starts with the extraction of the passing sequences, i.e., the order of players whom the ball traversed. Afterwards, we determine computationally the significance of the different k-pass-long motifs in the passing style of the teams. Our flow motif profile focuses on how ball traverses within a team. We not only count the number of passes, but also check which players are involved in, and how they organize the flow of passes. Based on these computed flow motif profiles, we finally cluster the teams. To the best of our knowledge, our study is the first of its kind that investigates motifs in soccer passing sequences. Our contribution is twofold:
\begin{enumerate}
	\item we propose a method to quantify the motif characteristics of soccer teams based on their pass networks, and 
	\item we identify similarities and disparities between teams and leagues using the teams' motif fingerprints.
\end{enumerate}

In the recent decade, several data-provider companies and websites have arisen to annotate soccer matches and to publish soccer datasets. For example, such initiatives include Prozone~\cite{prozone}, OptaPro~\cite{optapro}, Instat Football~\cite{instat}, and Squawka~\cite{squawka}, among others. The prevalence of data-providers enables us to take a data-driven, quantitative approach to identify the styles of the soccer teams. We focus on the 2012/13 seasons of major European soccer leagues and analyze the passing strategies of the teams throughout the whole season.

\section{Methodology}
The ``flow motifs'' of a pass network, in which players are linked via executed passes, consist of a given number of consecutive passes, namely, an ordered list of players who were involved in the particular passes. Throughout this paper we focus on motifs consisting of three consecutive passes, however, it is straightforward to generalize our methodology to investigate motifs with fewer/more passes. Our methodology relaxes the identity of the involved players, i.e., it does not differentiate motifs based on the names of the players, rather focuses on the certain structure of the passes. There are five distinct motifs when we analyze three-pass long motifs: ABAB, ABAC, ABCA, ABCB, and ABCD. For example, the motif ABAB denotes the following pass sequence: first, player 1 passes to player 2; second, player 2 passes the ball back to player 1; and finally, player 1 passes again to player 2. If a similar pass sequence happens between player 3 and player 4 the identified motif is ABAB again (i.e., the crucial characteristic is what happened and not between whom).

Our methodology quantifies the prevalence of the flow motifs in the pass networks compared to random networks whose degree distribution is the same. To achieve this, we start with a list of passes that a team made during a match. The format of a pass record is
\[
p_n = < \textrm{player}_i(n),\textrm{player}_j(n),t(n)>
\]
where $\textrm{player}_i(n)$ passed the ball to $\textrm{player}_j(n)$ in the $t(n)$ time instance. Second, we derive all the ball possessions that a team had. A ball possession $<p_1,p_2,⋯,p_n>$ consists of such passes that fulfill two constraints:
\begin{eqnarray*}
	\textrm{player}_j(m) = \textrm{player}_i(m+1), \forall m \in \{1,\dots,n-1\} \\
	t(m+1) - t(m) \leq T_{\textrm{max}}, \forall m \in \{1,\dots,n-1\}	
\end{eqnarray*}
where $T_{\textrm{max}}$ denotes the time threshold between two passes. These constraints assure that the passes are consecutive (i.e., a player receives the ball and then passes it forward) and not having major breaks. Throughout our study, we use $T_{\textrm{max}}=5\textrm{sec}$ to determine if two passes are belonging to the same ball possession.

Third, we extract all the three-pass long sub-possessions from the ball possessions (e.g., a ball possession having $n$ passes contains $n-2$ motifs) and convert the player identifiers into the appropriate A, B, C, and D labels to assemble the motifs. For example, a ball possession where the ball moves between players as $2 \rightarrow 4 \rightarrow 5 \rightarrow 6 \rightarrow 4 \rightarrow 6$ translates into three motifs, namely, ABCD, ABCA, and ABCB: 
\[\overunderbraces{&&\br{3}{ABCA}}%
{&2 \rightarrow &4 \rightarrow &5 \rightarrow 6& \rightarrow 4& \rightarrow 6&}
{&\br{3}{ABCD}&&}
\]

After having the motifs that are present in the pass network, we quantify the prevalence of the motifs by comparing the pass network of the team to random pass networks having identical properties (in particular, the number of vertices and their degree distribution). Specifically, we perturb the labels of the motifs prevalent in the original pass network randomly and such we create pseudo motif-distributions. In our data analyses, we generate 1000 random pass networks for each original pass network. Finally, we compute the z-scores (a.k.a. standard scores) of the motifs by comparing the original and the constructed random pass networks. As a result, we have a characteristic of the (passing) style of a team for every match---in terms of the z-scores of the motifs.

\begin{figure}[tb]
\centering
\includegraphics[width=9cm]{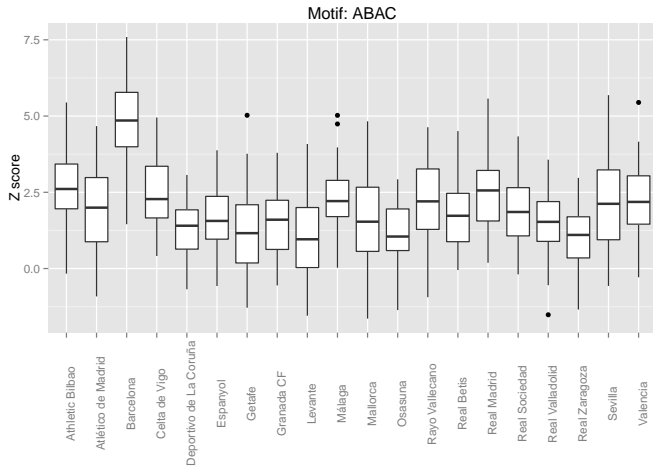}
\caption{The prevalence of the ABAC motif in case of the teams of the Spanish first division (median, quartiles) with respect to their z-scores. FC Barcelona applies the ABAC motif much more frequently than any other team in the league.}
\label{fig:spain_abac}
\end{figure}

\begin{figure}[tb]
\centering
\includegraphics[width=9cm]{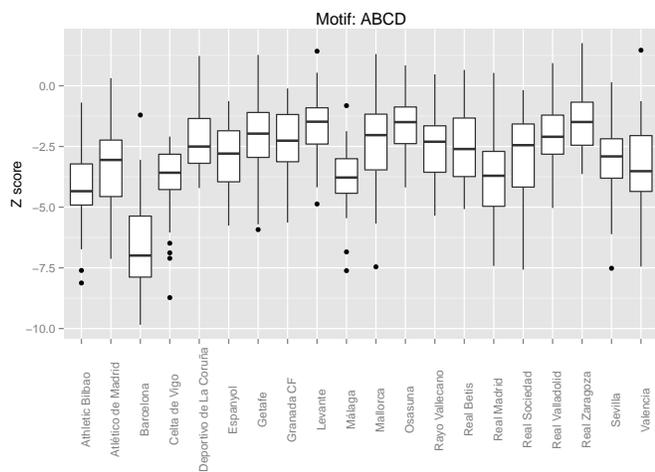}
\caption{FC Barcelona uses the ABCD motif less often than the other teams.}
\label{fig:spain_abcd}
\end{figure}

\begin{figure*}[tb]
\centering
\includegraphics[width=5.5cm]{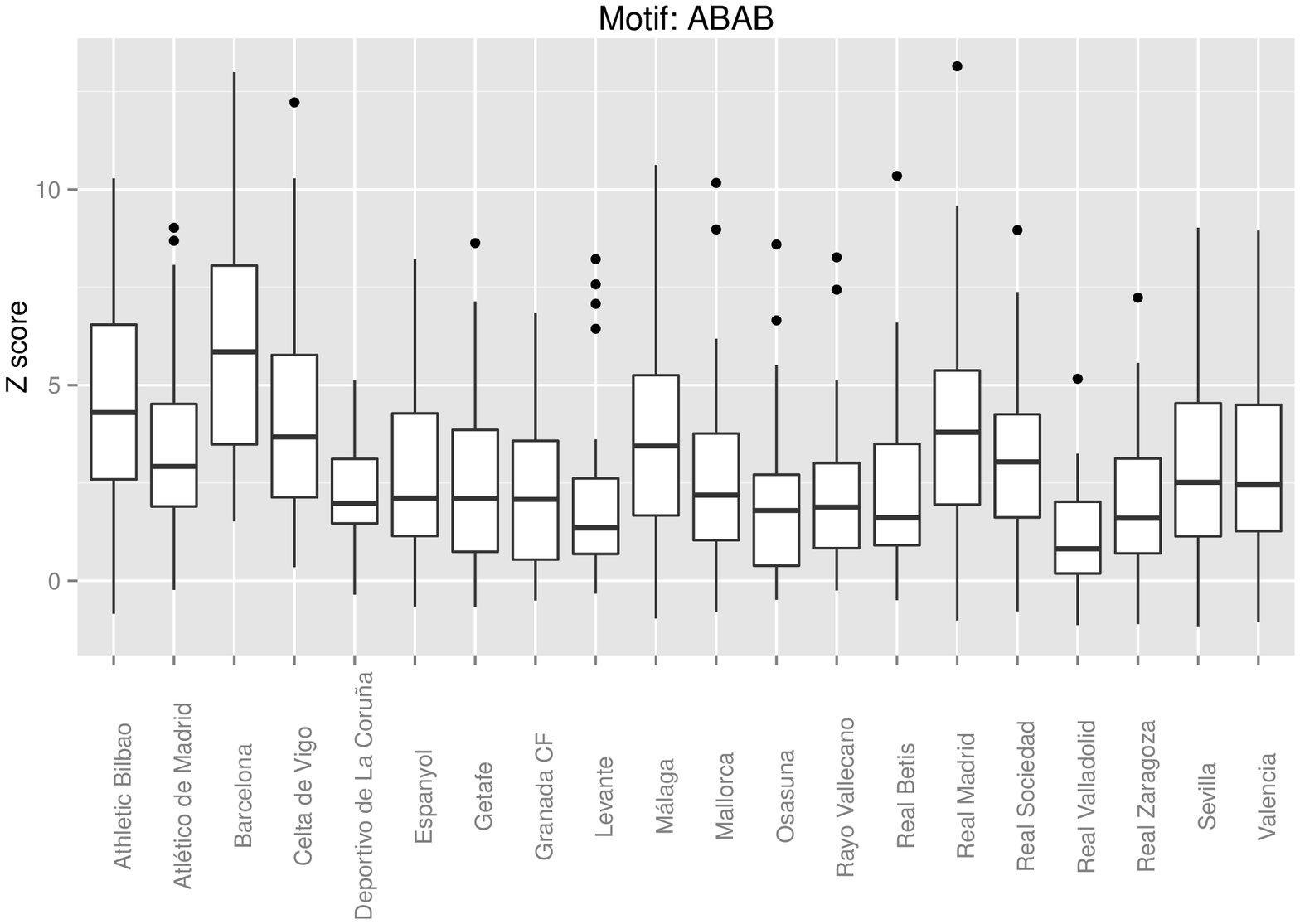}
\includegraphics[width=5.5cm]{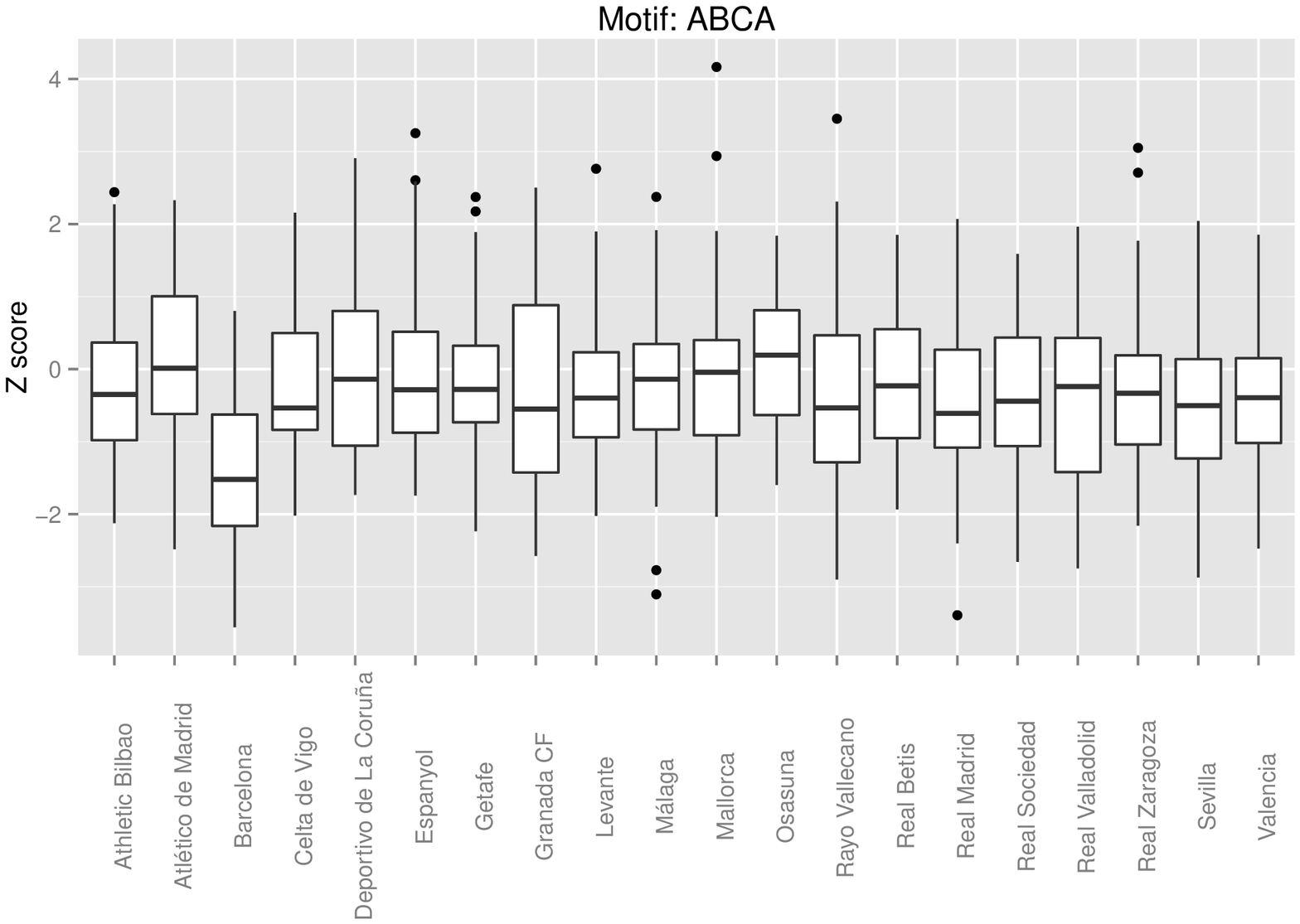}
\includegraphics[width=5.5cm]{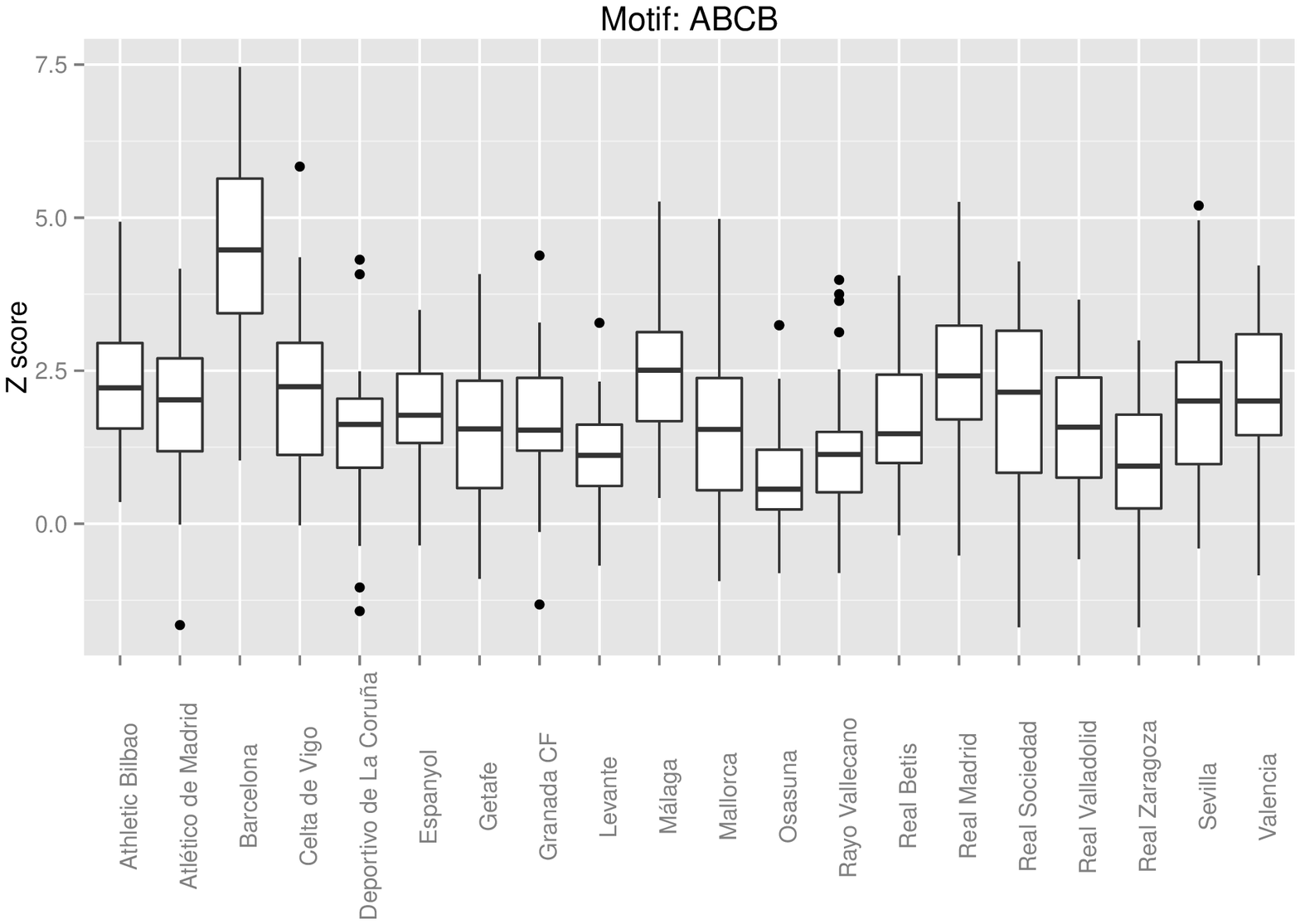}
\caption{Z-scores of the ABAB, ABCA, and ABCB motif in the Spanish league.}
\label{fig:spain_all}
\end{figure*}

\section{Data analysis and results}

We use publicly accessible information on the pass networks of soccer teams. In particular, the dataset contains information from the 2012/13 seasons of the Spanish, Italian, English, French, and German first division. For example, the part of the dataset that contains information on the Spanish league spreads 20 teams, 380 matches, and more than 250 thousands of passes. We quantify the motif characteristics of the teams using the aforementioned dataset. We first present results on the passing styles of teams in the Spanish first division and later on we compare our finding with the other European leagues and teams.

We compare the Spanish teams with respect to their ABAC motifs in Figure~\ref{fig:spain_abac}. Most of the teams have similar z-scores, i.e., apply the ABAC pass motif to comparable extent. However, FC Barcelona has a quite distinct strategy: applies ABAC motifs significantly more often than the other teams (the difference is at least 2.5 standard deviation). The trend is similar in case of the ABCD motif; the only difference is that the majority of the teams have notably larger z-scores than FC Barcelona (Figure~\ref{fig:spain_abcd}). This means that FC Barcelona applies this motif significantly less frequently than the other teams. In general, FC Barcelona uses structured motifs (i.e., motifs with more back and forth passes such as ABAB, ABAC, and ABCB) more often than simpler ones compared to other teams. We present the results of the remaining motifs in Figure~\ref{fig:spain_all}.

\begin{figure}[tb]
\centering
\includegraphics[width=9cm]{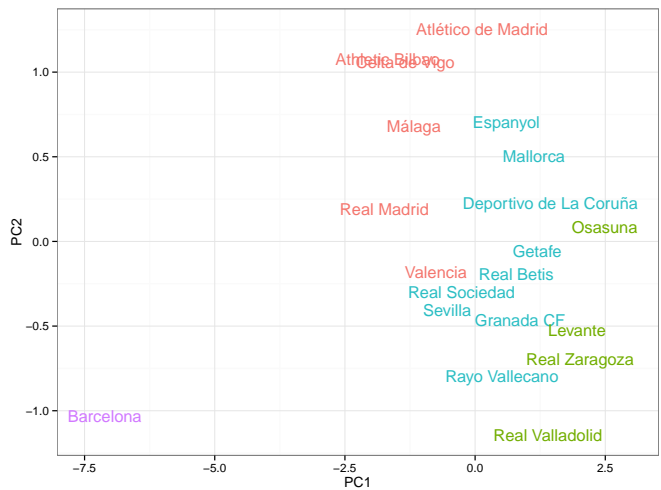}
\caption{K-means clustering of the teams in the Spanish league. One of the four clusters contains only a single team, namely, FC Barcelona that has an unique style based on its passing motifs.}
\label{fig:kmeans}
\end{figure}

\vspace{0.5cm}
We next analyze the similarities and the differences of the teams' motif characteristics via cluster analysis. First, for each team, we construct a feature vector representing the team's usage of motifs. We use the mean of the z-scores of the five distinct motifs as the features (by averaging the z-scores over 38 matches a team had in the season). Afterwards, we cluster the teams based on their five-motif long feature vectors. We use two methods for cluster analysis: k-means and hierarchical clustering. We illustrate the result of the k-means clustering in Figure~\ref{fig:kmeans} (the clusters are color-coded), where the ratio of the within the cluster and the total sum of squares is 90.3\%. For example, the cluster that contains Atletico Madrid and Athletic Bilbao, among others, is characterized by extensive usage of ABAB and ABCA motifs. While most of the teams are clustered in three major groups, FC Barcelona is separated from the other teams. FC Barcelona is the only team in its cluster; hence, it has a distinctive motif characteristics.

\begin{figure}[tb]
\centering
\includegraphics[clip=true,trim=1cm 2.5cm 0cm 2.2cm,width=9cm]{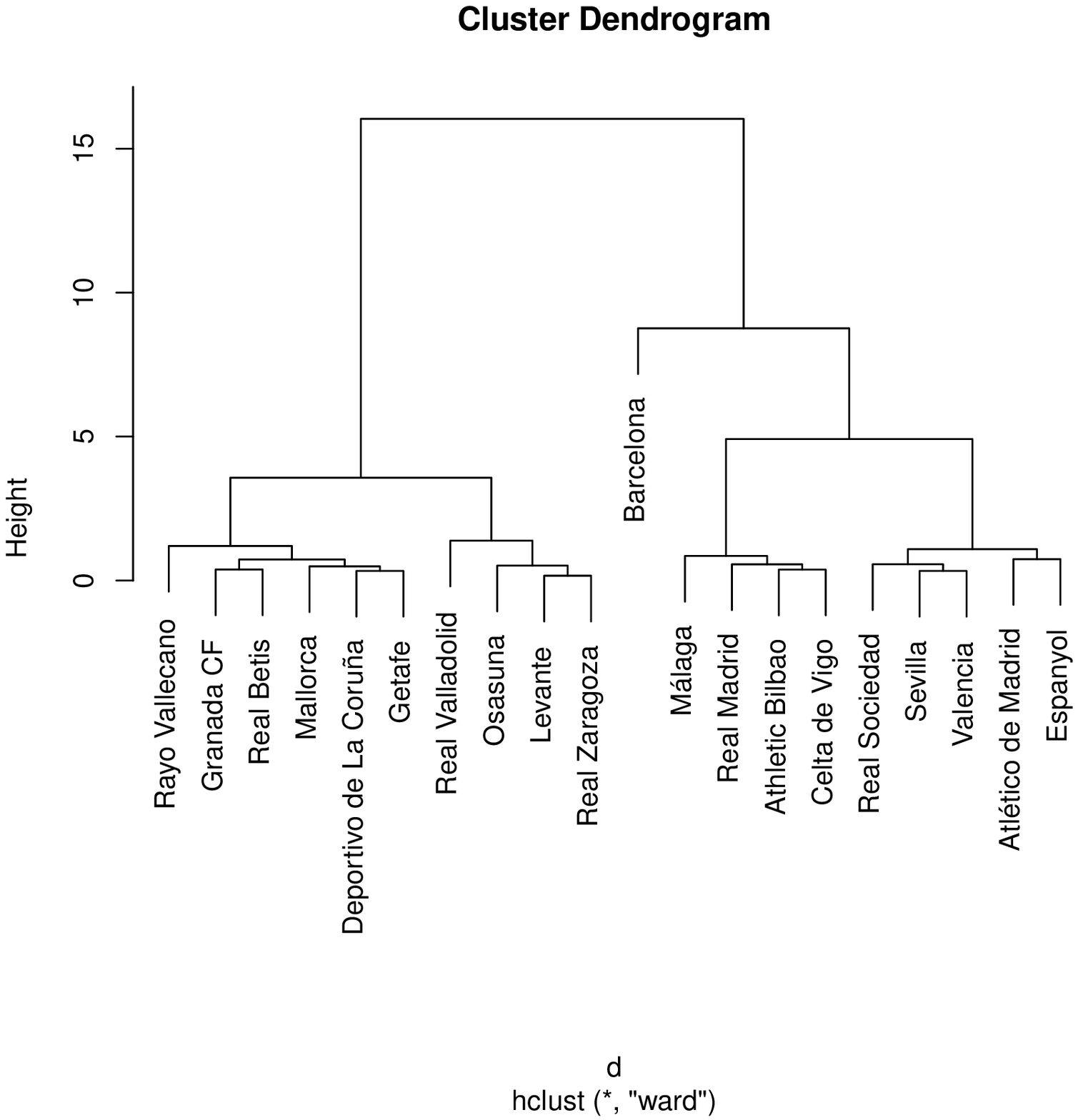}
\caption{Ward hierarchical clustering of the soccer teams in the Spanish league. FC Barcelona does not belong to any major groups of the teams.}
\label{fig:ward}
\end{figure}

The Ward hierarchical clustering algorithm reveals similar trend as shown in Figure~\ref{fig:ward}. Again, FC Barcelona has a solitary style while the other teams are having resembling features. The implications of the two clustering schemes are consistent: FC Barcelona had a unique, significantly different passing style than any other team in the Spanish league.

\begin{figure}[tb]
\centering
\includegraphics[width=8cm]{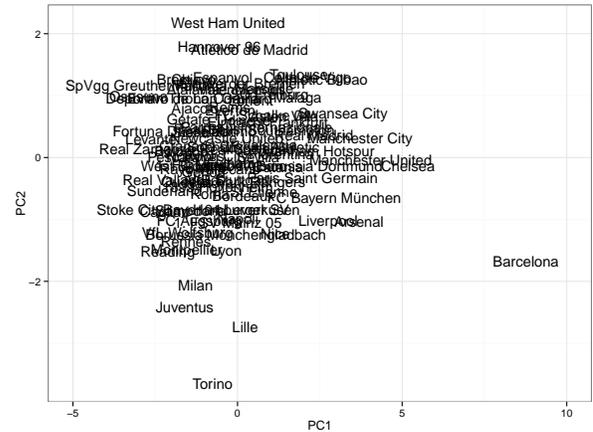}
\caption{The style of soccer teams of the Spanish, Italian, English, French, and German soccer leagues. FC Barcelona has a unique style even on a European scale.}
\label{fig:kmeans_all}
\end{figure}

Finally, we take a broader point of view and investigate whether the style of FC Barcelona remains unique if we consider teams of four additional European soccer leagues. We show the teams in Figure~\ref{fig:kmeans_all} based on their motifs using principal component analysis. Although we analyze more teams that have more variation in their pass characteristics, FC Barcelona still able to maintain its rare, distinct style. It is surprising that Torino, an Italian team nearly relegated at the end of the season, has a style diverse from the vast majority of the considered teams and shares properties with teams like Lille, Milan, and Juventus---dominant teams in the French and Italian leagues. The distinctive feature of Torino's strategy is that it involves less frequent usage of the ABCA motif.

\section{Future Work}
The presented results illustrate the potential of analyzing the flow motifs of soccer teams. There are several ways to extend the investigation of pass motifs to reveal finer-grained details of teams and players. As future work, we plan to address three areas: (i) condition the pass motifs based on the results of the matches, (ii) study the impact of home and away games on the prevalence of the motifs, and (iii) explore the players' involvement in the different motifs.

\section{Conclusions}
We proposed a quantitative method to evaluate the styles of soccer teams through their passing structures. The analysis of the motifs in the pass networks allows us to compare and differentiate the styles of different teams. Although most teams tend to apply homogenous style, surprisingly, a unique strategy of soccer is also viable---and quite successful, as we have seen in the recent years. Our results shed light on the unique philosophy of FC Barcelona quantitatively: the famous tiki-taka does not consist of uncountable random passes but rather has a precise, finely constructed structure.


\newpage
%
\bibliographystyle{abbrv}
\bibliography{pass_motifs_ref}  
%
%

%
%
%

%
%
%
%

\end{document}